\documentclass[twocolumn,aps,prl]{revtex4}
\def\be{\begin{equation}}
\def\ee{\end{equation}}
\def\bea{\begin{eqnarray}}
\def\eea{\end{eqnarray}}
\def\ba{\begin{array}}
\def\ea{\end{array}}

\def\d{\delta}

\def\0{$\Gamma_0$}
\def\o{\omega}

\def\t{\theta}

\begin{document}
%\draft
\title{Comment on ``Wave Refraction in Negative-Index Materials:
Always Positive and Very Inhomogeneous"}
\maketitle

%\begin{multicols}{2}
%\narrowtext

Valanju, Walser and Valanju (VWV) \cite{Valanju02} have shown that for 
a group consisting of two 
plane waves incident on the interface between a material of positive
refractive index (PIM) and material of negative refractive index (NIM), 
the
group velocity refracts positively. Here we show that this is true only 
for
the special two plane wave case constructed by VWV, but for generic
localized wave packets, the group refraction is generically negative.

The sum of two plane waves of wavevector and frequency $({\bf 
k}_1,\omega_1)$ and $({\bf k}_2,\omega_2)$, considered by VWV, can be 
written as 
$2e^{i({\bf k}_0\cdot {\bf r}-\omega_0 t)}
\cos[(1/2)({\bf \Delta k}\cdot {\bf r}-
\Delta\omega t)]$,
where $({\bf k}_0,\omega_0)$ the average wavevector 
and frequency and $({\bf \Delta k},\Delta\omega)$ denote their 
differences. 
Clearly, the argument of the cosine is constant along planes, which 
propagate in time along 
the 
direction of their normal, ${\bf\Delta k}$. We have carried out
numerical simulations of wave packets incident on the PIM-NIM interface 
and
for the 2-wave case arrive at conclusions similar to VWV. For arbitrary 
number of incident plane waves whose ${\bf k}$ vectors are all 
parallel, 
the group
refraction is again positive. Note that in all these special cases the
packet remains nonzero on infinite planes.

Here we show that for any wave packet that is spatially localized,
the group refraction is {\em generically negative}. 

For 3 (or more) waves whose wave vectors not aligned, the group 
refraction
will be negative. For example, consider three wave vectors in PIM in 
the $x$-$z$-plane, 
whose magnitudes are, $k-\d k$, $k$, $k-\d k$ and whose angles with the 
$z$-axis are, 
$\t-\d\t$, $\t$, $\t+\d\t$, respectively. The dispersion 
$k^2=(\o^2-\o_p^2)(\o^2-\o_b^2)/c^2(\o^2-\o_0^2)$ were used for NIM.
The results are shown in Fig. 1. The wave packet refracts negatively, 
in obvious
contrast to VWV.  As we have seen, 
two plane waves result in a wave packet-like structure which is 
constant 
along planes; the addition of a third wave 
breaks the planes into localized wave packets which 
refract negatively.

A packet constructed from a finite number of plane waves will always 
give 
a collection of propagating wave pulses, as seen in Fig. 1. A wave 
packet 
localized in one region of space, as occurs in all 
experimental situations, can only be constructed from a continuous 
distribution of wavevectors. Consider such a wave packet incident from 
outside the NIM,
$E=E_0\int d^2 k f({\bf k}-{\bf k}_0) 
e^{i({\bf k}\cdot {\bf r}-\omega ({\bf k})t)}$,
where $\omega({\bf k})=ck$. If $f({\bf k}-{\bf k}_0)$ drops off rapidly 
as 
{\bf k} moves away from ${\bf k}_0$, $\omega ({\bf k})$ can be expanded 
in a 
Taylor series to first order in ${\bf k}-{\bf k}_0$ to a good 
approximation. 
This gives,
$E=E_0e^{i({\bf k}_0\cdot {\bf r}-\omega ({\bf k}_0)t}
g({\bf r}-ct{\bf k}_0/k_0)$,
where $g({\bf R})=\int d^2 k f({\bf k}-{\bf k}_0)
e^{i({\bf k}-{\bf k}_0)\cdot {\bf R}}$.
Inside the NIM, {\bf k} and ${\bf k}_0$ in the argument of the 
exponent get replaced by ${\bf k}_r$ and ${\bf k}_{r0}$ which are 
related to {\bf k} and ${\bf k}_0$ by Snell's law. Then the wave packet 
once it enters the NIM is given by
\be
E_r=E_0'e^{i({\bf k_{r0}}\cdot {\bf r}-\omega ({\bf k_{r0}})t}
g_r({\bf r}-{\bf v}_{gr}t),
\ee
where 
$g_r({\bf R})=\int d^2 k f({\bf k}-{\bf k_0})
e^{i{\bf R}
\cdot [({\bf k}-{\bf k_0})\cdot
\nabla_{\bf k}{\bf k_r}]}$.
Here ${\bf k}_{r0}$ denotes ${\bf k}_{r}$ evaluated at ${\bf k}={\bf 
k}_{0}$ 
and ${\bf v}_{gr}=\nabla_{{\bf k}_r}\omega ({\bf k}_{r})$ evaluated at 
${\bf k}_r={\bf k}_{r0}$. 
Thus, the refracted wave moves with the group velocity ${\bf v}_{gr}$. 
Evaluation of Eq. (1) for a Gaussian wave packet shows that the 
incident 
packet gets distorted but the maximum of the packet moves at 
${\bf v}_{gr}$.  For the case 
of an isotropic medium, considered by VWV \cite{Valanju02}, the group 
velocity is anti-parallel to the wave vector in 
the 
medium. Hence, the group velocity will be 
refracted the same way as the wavevector is, contrary to the claim of 
VWV \cite{Valanju02}.

Thus VWV's statement that the ``Group Refraction is always positive'' 
is
true only for the very special (and rare) wave packets constructed by 
them
and is incorrect for more general wave packets that are spatially 
localized.

This work was supported by the National Science Foundation, the Air 
Force
Research Laboratories and the Department of Energy.

W. T. Lu, J. B. Sokoloff and S. Sridhar

Department of Physics, Northeastern University, 360 Huntington Avenue,
Boston, MA 02115.

\begin{figure}[htbp]
%\vskip 0.2cm
%\centereps{3.8in}{4.2in}{NIM_fig1.eps}
%\center{\special{eps: NIM_fig1-new.eps x=4in y=5in}}
%\vskip -1.8in
\caption{Negative refraction of 3 plane waves with $k=6.32$, $\d 
k=0.32$, $\t=\pi/4$, $\d\t=\pi/60$, $\o_0=2$, $\o_b=8$, $\o_p=10$, and $c=1$. 
(Inset above) Wavevectors for 
the 3 plane waves in the PIM (left) and NIM (right). The thick arrows indicate the 
wave packet propagation direction.}
\label{Fig1}
\end{figure}

%\end{multicols}{2}

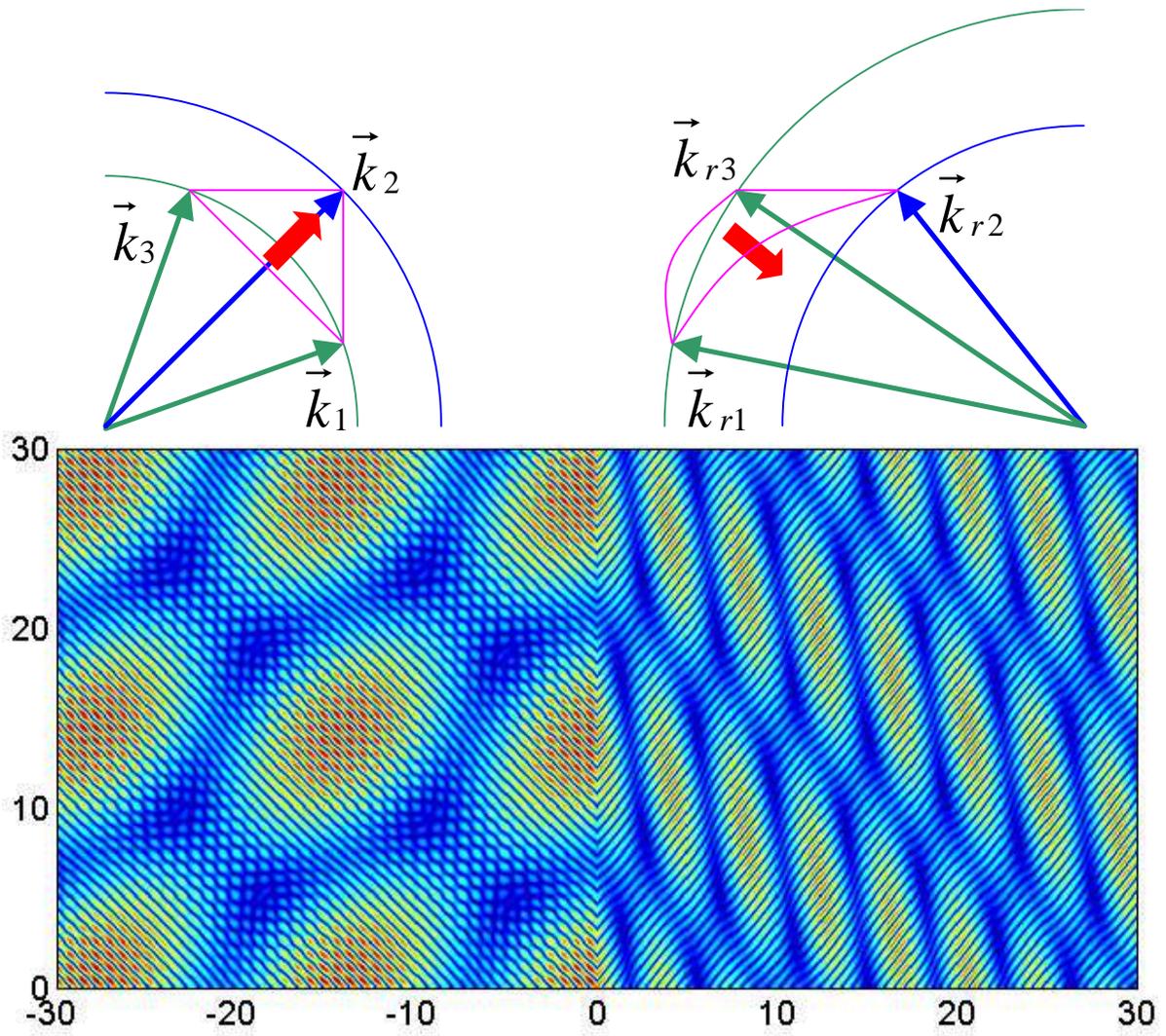
\end{document}